\documentclass[a4paper,11pt]{article}
\usepackage[latin1]{inputenc}
\usepackage[english]{babel}
\usepackage{graphicx}
\begin{document}
\title{A coupled vacuum energy model producing endless alternated phases of accelerated and decelerated expansion}
\author{Stéphane Fay\footnote{steph.fay@gmail.com}\\
Palais de la Découverte\\
Astronomy Department\\
Avenue Franklin Roosevelt\\
75008 Paris\\
France}
\maketitle
\begin{abstract}
We consider a flat Universe filled with a vacuum energy coupled to matter and radiation by respectively positive coupling functions $Q_m$ and $Q_r$. We require that these functions be such as Universe exits from inflation to go to a radiation dominated epoch and allow to reproduce the observed $\Lambda CDM$ expansion after this last epoch. These requirements lead to some necessary constraints on the coupling functions. We then look at one of the simplest forms of $Q_m$ and $Q_r$ able to satisfy them. The cosmological model thus defined describes a Universe with an initial singularity and endless alternated periods of accelerated and decelerated expansion (one of them being our $\Lambda CDM$ Universe) unifying inflation and present time expansion acceleration.
\end{abstract}
\section{Introduction} \label{s0}
We consider a flat Universe filled with a vacuum energy\cite{Car01} coupled to matter and radiation\cite{Lim13} by respectively positive coupling functions $Q_m$ and $Q_r$. Couplings between dark energy and other species\cite{Boh10} have been used even before the dark energy discovery\cite{Per99, Rie98}. They allowed to study various problems such as dark matter with varying mass\cite{Gar93} or the cosmological constant problem\cite{Wet95}. Some observational constraints on these coupled models have been proposed, mainly when radiation is negligible\cite{Ame00,Oli05,Yan14,Cos14}, based on supernovae, cosmological background radiation (CMB) and density perturbations. Coupling between dark energy and radiation has been specifically studied from the viewpoints of thermodynamics\cite{Gas87, Pav88, Lim96}, minimal noncanonical cosmologies\cite{Bar06}, flux destabilisation\cite{Hor07} or inflaton decay to radiation\cite{Suy08,Her10}. In this paper we present a cosmological model able to describe the whole Universe history\cite{Lim13,Per13,Lim15} by choosing $Q_m$ and $Q_r$ on the base of two physical constraints.\\
The first constraint is that $Q_m$ and $Q_r$ allow an exit from inflation\cite{Gut81,Bra01,Lin04,Bic14,Ada14} to go to a radiation dominated Universe. Since $Q_m$ is positive, some vacuum energy is continuously cast into matter. To get a radiation dominated Universe after inflation, we thus have to control that matter stays negligible with respect to radiation\cite{Bea01,Fre87} until radiation dominated epoch. This requirement leads to a lower limit on the ratio $Q_r/Q_m$. The second constraint consists in finding some necessary conditions on $Q_m$ and $Q_r$ such as vacuum energy leads to a $\Lambda CDM$ expansion after the radiation dominated epoch. Then, one of the simplest forms of $Q_m$ and $Q_r$ respecting these two constraints describes a Universe with an initial singularity and endless alternated phases of accelerated and decelerated expansion unifying inflation and present day expansion acceleration.\\
The plan of the paper is as follows. In section \ref{s1}, we write the field equations under the form of a dynamical system. In section \ref{s2}, we look for some conditions on $Q_m$ and $Q_r$ necessary to satisfy the two above mentioned constraints. In section \ref{s3}, we chose one of the simplest forms of $Q_m$ and $Q_r$ able to satisfy them. We then get a cosmological model with endless alternated phases of accelerated and decelerated expansion. We check that it is in agreement with some supernovae data and study its properties (age, periodicity, $\Lambda CDM$ approximation, etc). We conclude in the last section.
\section{Field equations} \label{s1}
\begin{figure}[h]
\centering
\includegraphics[width=6cm]{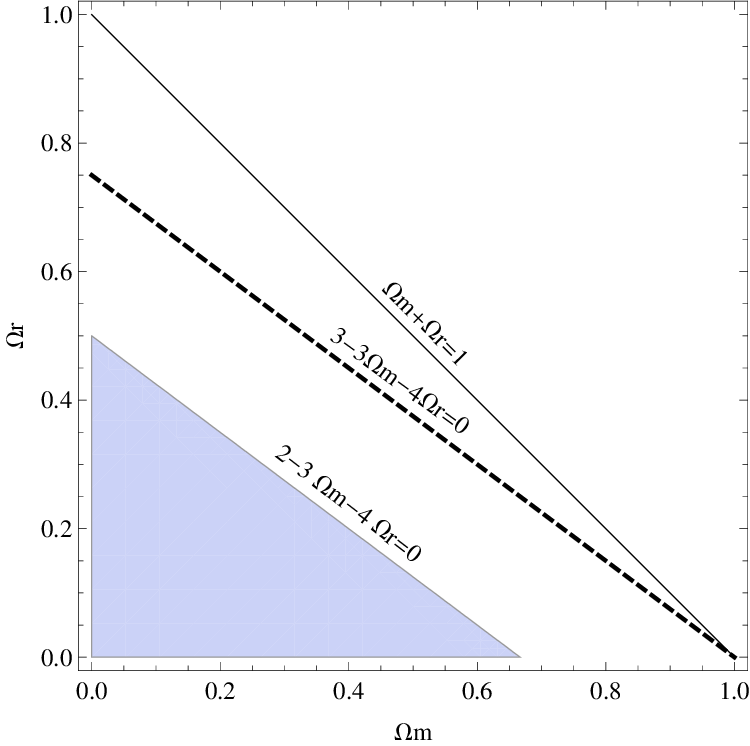}
\caption{\scriptsize{\label{fig0}Solutions to the field equations are such that $\Omega_m+\Omega_r\leq 1$. The accelerated expansion takes place in the gray area below the line $2-3\Omega_m-4\Omega_r=0$. Above the dashed line $3-3\Omega_m-4\Omega_r=0$, the slope $s$ of a trajectory is always such that $s<q_r/q_m$ and $\Omega_m'>0$.}}
\end{figure}

The field equations of General Relativity for a flat homogeneous and isotropic Universe filled with a dark energy coupled to matter and radiation write
\begin{equation}\label{H2}
H^2=\frac{k}{3}(\rho_m+\rho_r+\rho_d)
\end{equation}
\begin{equation}\label{rhomd}
\dot \rho_m+3H\rho_m=Q_m
\end{equation}
\begin{equation}\label{rhord}
\dot\rho_r+4H\rho_r=Q_r
\end{equation}
\begin{equation}\label{rhodd}
\dot\rho_d+3(1+w)H\rho_d=-Q_m-Q_r
\end{equation}
$H$ is the Hubble function and $w$, the dark energy equation of state. $\rho_m$, $\rho_r$ and $\rho_d$ are respectively the densities of matter, radiation and dark energy. We choose this last one as a vacuum energy with $w=-1$. A dot means a derivative with respect to proper time $t$. We assume that Universe is expanding, i.e. $H>0$, as indicated by observations. We also assume that the densities and coupling functions $Q_i$ are positive. As noted in \cite{Cam15}, positive coupling functions generally allow to alleviate the coincidence problem. In \cite{Mur16} it has also been found that when dark matter is fed by dark energy, the independent determinations of the Hubble constant and the amplitude of the linear power spectrum $\sigma_8$ agree for high and low redshift observations. In the opposite case, when dark matter is cast into dark energy, there are then some tensions between the determinations of these parameters at high and low redshift. Note also that in \cite{Cam15}, defining $r_m=\rho_m/\rho_d$ and $r_r=\rho_r/\rho_d$, it is found that the positivity of the $Q_i$ implies that $(\dot r_m/r_m,\dot r_r/r_r)<3Hw$, restricting the rate at which $r_m$ and $r_r$ could decrease with expansion.\\
To rewrite the field equations, we define the dimensionless variables
\begin{equation}\label{y1}
\Omega_m=\frac{k}{3}\frac{\rho_m}{H^2}
\end{equation}
\begin{equation}\label{y2}
\Omega_r=\frac{k}{3}\frac{\rho_r}{H^2}
\end{equation}
\begin{equation}
\Omega_d=\frac{k}{3}\frac{\rho_d}{H^2}
\end{equation}
\begin{equation}\label{q1d}
q_m=\frac{k}{3}\frac{Q_m}{H^3}
\end{equation}
\begin{equation}\label{q2d}
q_r=\frac{k}{3}\frac{Q_r}{H^3}
\end{equation}
The present values of the density parameters $\Omega_m$, $\Omega_r$ and $\Omega_d$ are noted as $\Omega_{m0}$ for matter, $\Omega_{r0}$ for radiation and $\Omega_{d0}=1-\Omega_{m0}-\Omega_{r0}$ for vacuum energy. This last equality comes from equation (\ref{H2}) that gives the constraint
\begin{equation}\label{cons}
\Omega_m+\Omega_r+\Omega_d=1
\end{equation}
With the variables (\ref{y1})-(\ref{q2d}), the field equations rewrite
\begin{equation}\label{eq1}
\Omega_m'=\Omega_m(3\Omega_m+4\Omega_r-3)+q_m
\end{equation}
\begin{equation}\label{eq2}
\Omega_r'=\Omega_r(3\Omega_m+4\Omega_r-4)+q_r
\end{equation}
A prime means a derivative with respect to $N=\ln a$, $a$ being the scale factor of the FLRW metric. We choose at present time $a=1$ and thus $N=0$. The $q_i$ and $\Omega_i$ being, at least formally, some functions of the redshift $z$, the $q_i$ can also be considered as some functions of the $\Omega_i$. The dynamical system (\ref{eq1}-\ref{eq2}) is thus autonomous. For instance, when $q_i\propto \Omega_i$, we recover some classical forms of the coupling functions $Q_i\propto H \rho_i$ \cite{Boe08, Oli06}. Since the Hubble function $H>0$, $N$ is an increasing function of the proper time $t$ and thus a time variable. Moreover, the assumption that energy densities are positive implies that $(\Omega_m,\Omega_r,\Omega_d)>0$. It then follows from the constraint (\ref{cons}) that the density parameters are also normalised i.e $(\Omega_m,\Omega_r,\Omega_d)<1$. The solutions to the field equations can thus be represented as trajectories in a finite part  $\Omega_m+\Omega_r<1$ of the phase space $(\Omega_m,\Omega_r)$. This part is plotted on figure \ref{fig0}.\\
It contains a set of points $(\Omega_m,\Omega_r)$ for which Universe expansion is accelerated, i.e. $d^2a/dt^2>0$. This second derivative has the same sign as $\frac{\dot H}{H^2}+1$. This set of points is thus defined by:
\begin{equation}\label{acc}
2-3\Omega_m-4\Omega_r>0
\end{equation}
It is independent from the coupling functions $Q_i$ and is plotted in gray on figure \ref{fig0}. 
Finally, we define the slope $s=d\Omega_r/d\Omega_m$ of a trajectory as
\begin{equation}\label{slope}
s=\frac{\Omega_r(3\Omega_m+4\Omega_r-4)+q_r}{\Omega_m(3\Omega_m+4\Omega_r-3)+q_m}
\end{equation}
\section{Physical constraints on the coupling functions} \label{s2}
We want to choose $Q_m$ and $Q_r$ on the bases of two physical constraints. The first one is that after inflation, Universe becomes radiation dominated. The second one is that vacuum energy leads to a $\Lambda CDM$ expansion after the radiation dominated epoch, in agreement with observations. In this section, we derive some conditions on the coupling functions that are necessary to satisfy these two requirements.
\subsection{A necessary condition on $q_r/q_m$ to reach a radiation dominated epoch after inflation}\label{s21}
The positive coupling between vacuum and matter means that vacuum is continuously cast into matter. If, after exciting from inflation, we want that Universe be radiation dominated such as $(\Omega_m,\Omega_r)\rightarrow (\Omega_{m(dom)}<<1,\Omega_{r(dom)}\simeq 1)$, we thus have to restrain the increase of the matter density parameter with respect to this of the radiation. We are going to show that it implies a lower limit on $q_r/q_m$ as a function of $\Omega_{m(dom)}$ and $\Omega_{r(dom)}$. To calculate this limit, we start by considering the following inequalities:
\begin{itemize}
\item $s>q_r/q_m$
\item $s>0$ when $\Omega_r'<0$
\item $s<0$ when $\Omega_r'>0$
\end{itemize}
All these inequalities require the same necessary condition to be respected:
\begin{equation}\label{lim}
\Omega_r<\frac{3}{4}(1-\Omega_m)
\end{equation}
The line defined by (\ref{lim}) is plotted on figure \ref{fig0}. This figure shows that a trajectory going from inflation to radiation dominated epoch necessarily crosses the line (\ref{lim}) once or several times. More specifically, we are interested by the part of this trajectory between its last crossing of the line (\ref{lim}) and the radiation dominated epoch in $(\Omega_{m(dom)},\Omega_{r(dom)})$. We call this part $\Pi$. On $\Pi$, we thus have that the slope $s<q_r/q_m$. Moreover, we also have that $s=d\Omega_r/d\Omega_m$ has the same sign as $\Omega_r'$. Hence, $\Omega_m'>0$ and $\Omega_m$ is always increasing on $\Pi$\footnote{Remark that $Q_m>0$ means that dark energy is cast into matter not necessarily that $\rho_d'$ and $\Omega_m'$ are positive.}.\\
This last inequality implies that a trajectory reaching $(\Omega_{m(dom)},\Omega_{r(dom)})$ must be such as $\Omega_m<\Omega_{m(dom)}$ on $\Pi$. Otherwise, $\Omega_m$ would be larger than $\Omega_{m(dom)}$ and the radiation dominated epoch in $(\Omega_{m(dom)},\Omega_{r(dom)})$ could not be reached. Physically, this inequality is a constraint on the quantity of vacuum energy that can be cast into dark matter on $\Pi$. Note that such a constraint does not apply to radiation. If $\Omega_r$ becomes larger than $\Omega_{r(dom)}$ on $\Pi$, it can decrease to reach this last value with $s<0$. This is thus the exchange between vacuum energy and matter that must be controlled on $\Pi$.\\
Now let us assume that $\Pi$ is a straight line. Then, its smallest possible slope is such as $\Pi$ pass through the point having the largest value for $\Omega_r$ on the curve (\ref{lim}), i.e. $\Omega_r=3/4$ when $\Omega_m=0$, and go to $(\Omega_{m(dom)},\Omega_{r(dom)})$. This slope is thus
$$
s_{dom}=(\Omega_{r(dom)}-3/4)\Omega_{m(dom)}^{-1}
$$
$s_{dom}$ is very large since $(\Omega_{m(dom)}<<1,\Omega_{r(dom)}\simeq 1)$. Any other straight line $\Pi$ leaving (\ref{lim}) to reach $(\Omega_{m(dom)},\Omega_{r(dom)})$ will have a slope larger than $s_{dom}$. Any other curve $\Pi$ that is not a straight line will have some parts of it with $s<s_{dom}$ and the other parts such as $s>s_{dom}$. In particular, such a curve $\Pi$ cannot reach the point $(\Omega_{m(dom)},\Omega_{r(dom)})$ by zigzagging in the $(\Omega_m,\Omega_r)$ plane, thus keeping its slope always such as $s<s_{dom}$, since $\Omega_m'>0$. It thus follows that the maximum slope $s_{max}$ on $\Pi$ is always such as 
\begin{equation}\label{ineq1}
s_{max}\geq s_{dom}
\end{equation}
We thus always have $s>s_{dom}$ during one or several finite periods of times on $\Pi$ (the precise determination of these periods depends on the forms of the coupling functions). But we also know that above the curve (\ref{lim}), the slope $s$ of a trajectory is always smaller than $q_r/q_m=Q_r/Q_m$. Consequently, a trajectory reaches the radiation dominated epoch in $(\Omega_{m(dom)},\Omega_{r(dom)})$ if the necessary condition
\begin{equation}\label{ineq2}
q_r/q_m\geq s_{dom}
\end{equation}
is respected during the above mentioned finite periods of time. If it is not respected, matter increases too much and, on such a trajectory, the radiation dominated epoch in $(\Omega_{m(dom)},\Omega_{r(dom)})$ cannot be reached after inflation. Hence, $s_{dom}$ being large, the condition (\ref{ineq2}) cannot be respected if $q_r\simeq q_m$. In contrast, one of the simplest forms of $q_i$ able to fill the condition (\ref{ineq2}) is $q_r=\alpha$ with $\alpha$ a constant and $q_m=\Omega_m$ (since in this paper we consider $q_m\not =0$).
\subsection{A necessary condition on $q_r$ and $q_m$ such as vacuum energy leads to a $\Lambda CDM$ expansion}\label{s22}
\begin{figure}[h]
\centering
\includegraphics[width=6cm]{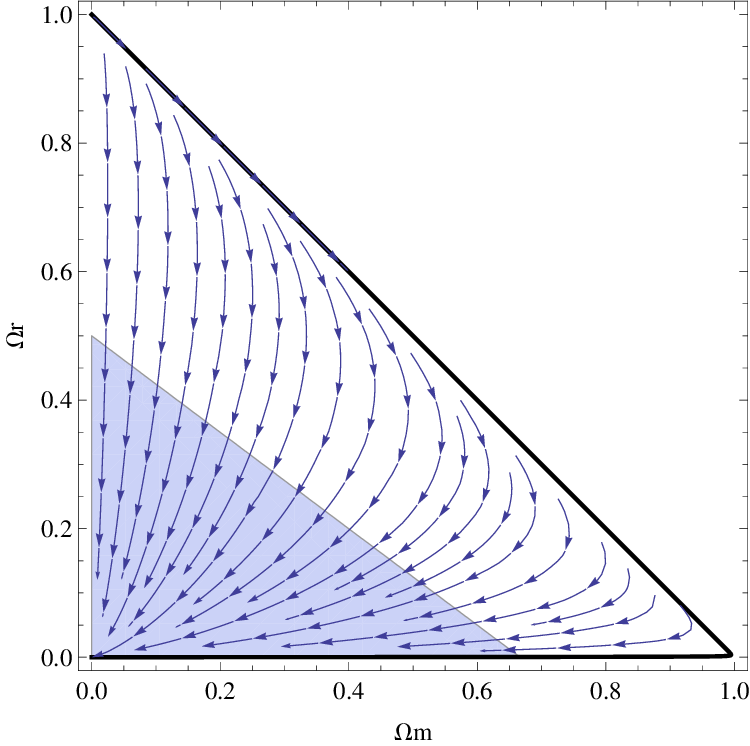}
\caption{\scriptsize{\label{fig1}Phase space of the $\Lambda CDM$ model. The thick trajectory corresponds to our Universe with at present time $\Omega_{m0}=0.27$ and $\Omega_{r0}=8.27\times 10^{-5}$.}}
\end{figure}

We want that vacuum energy leads to a $\Lambda CDM$ expansion after the radiation dominated epoch. It means that the coupling functions $Q_m$ and $Q_r$ should be negligible\footnote{Remark that they can be vanishing but non negligible or diverging but negligible in the fields equations.} in the energy conservation equations (\ref{rhomd}-\ref{rhodd}) or $q_m$ and $q_r$ in equations (\ref{eq1}-\ref{eq2}). For this last requirement, a necessary condition is that $q_m$ and $q_r<<1$. To define more precisely these last inequalities, we plot the phase space of the $\Lambda CDM$ model on figure \ref{fig1}.
We remark that after the radiation dominated phase, the trajectory corresponding to our Universe with at present time $\Omega_{m0}=0.27$ and $\Omega_{r0}=8.27\times 10^{-5}$ follows closely the lines $\Omega_d=0$ and then $\Omega_r=0$. It is plotted as a thick trajectory on figure \ref{fig1}. It is thus necessary that $q_r$ and $q_m<<1$ along these lines such as Universe expansion could behave like the one of a $\Lambda CDM$ model in agreement with some observations such as supernovae.
\section{A cosmological model with endless alternated phases of accelerated and decelerated expansion} \label{s3}
\begin{figure}[h]
\centering
\includegraphics[width=6cm]{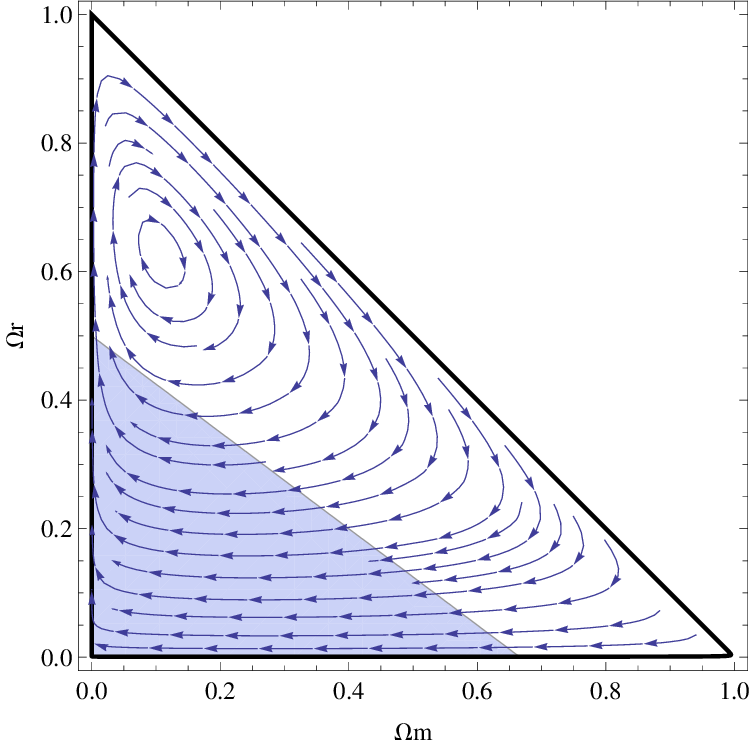}
\caption{\scriptsize{\label{fig2}Phase space of the model defined by $q_m=\Omega_m\Omega_r\Omega_d$ and $q_r=\alpha\Omega_r\Omega_d$ with $\alpha=4.4$. The thick trajectory corresponds to our Universe with today $\Omega_{m0}=0.27$ and $\Omega_{r0}=5.78\times 10^{-4}$.}}
\end{figure}

\begin{figure}[h]
\centering
\includegraphics[width=12cm]{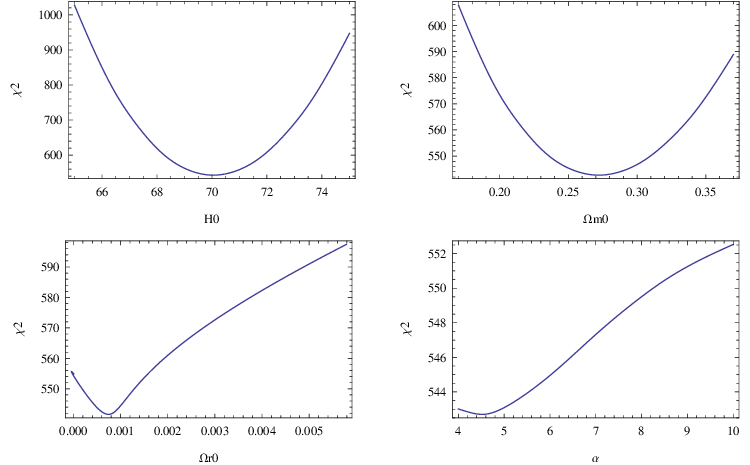}
\caption{\scriptsize{\label{fig7}Evolution of the $\chi^2$ with respect to $H_0$, $\Omega_{m0}$, $\Omega_{r0}$ and $\alpha$.}}
\end{figure}

\begin{figure}[h]
\centering
\includegraphics[width=6cm]{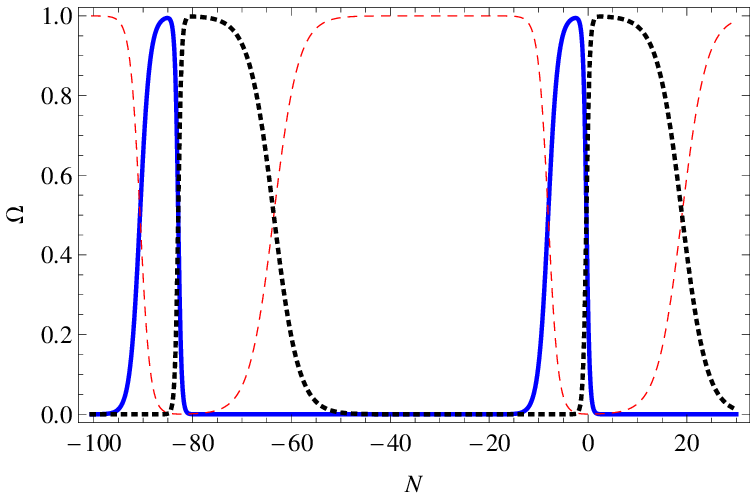}
\includegraphics[width=6cm]{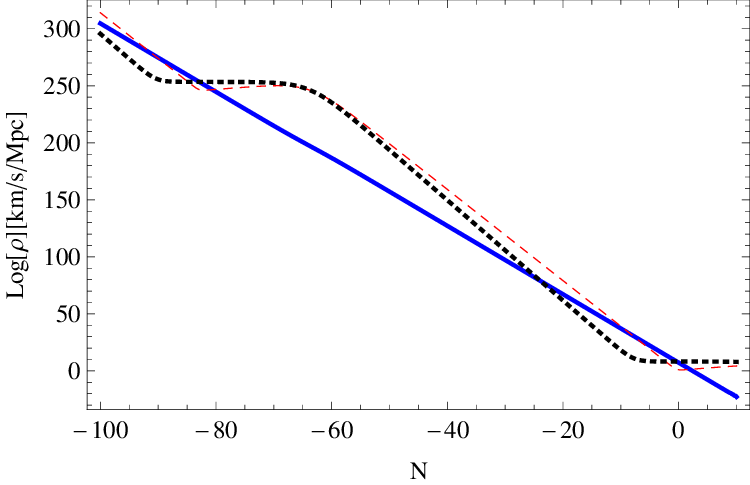}
\caption{\scriptsize{\label{fig3}Evolution of the density parameters (first graph) and densities (second graph) of matter (thick), radiation (dashed) and vacuum (thick dotted) with units defined as $k/3=1$.}}
\end{figure}

\begin{figure}[h]
\centering
\includegraphics[width=6cm]{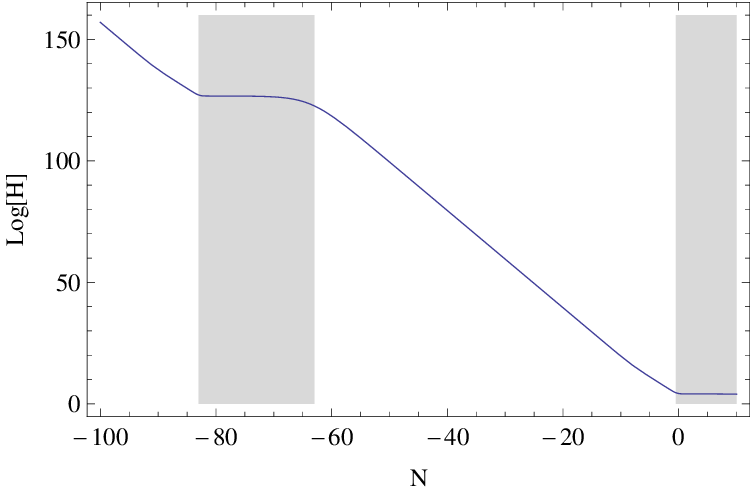}
\includegraphics[width=6cm]{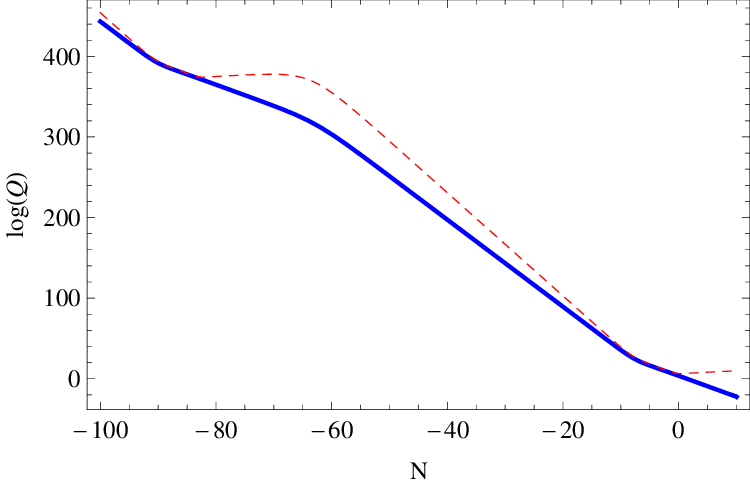}
\caption{\scriptsize{\label{fig4}The Hubble function (first graph with gray areas representing periods of times when expansion is accelerated) and the coupling functions (second graph) $Q_m$ (thick) and $Q_r$ (dashed).}}
\end{figure}

\begin{figure}[h]
\centering
\includegraphics[width=6cm]{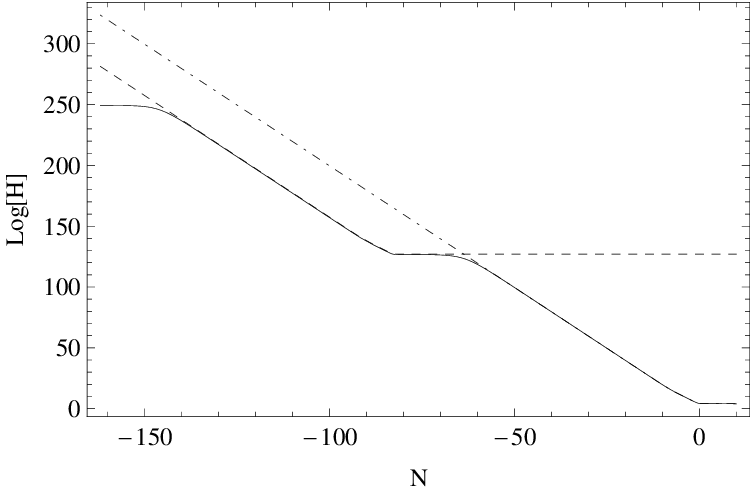}
\caption{\scriptsize{\label{fig5}The Hubble function (thin) can be successively approximated by some $\Lambda CDM$ models (dashed and dashed/dotted curves) but during the transition between two epochs.}}
\end{figure}

\begin{figure}[h]
\centering
\includegraphics[width=6cm]{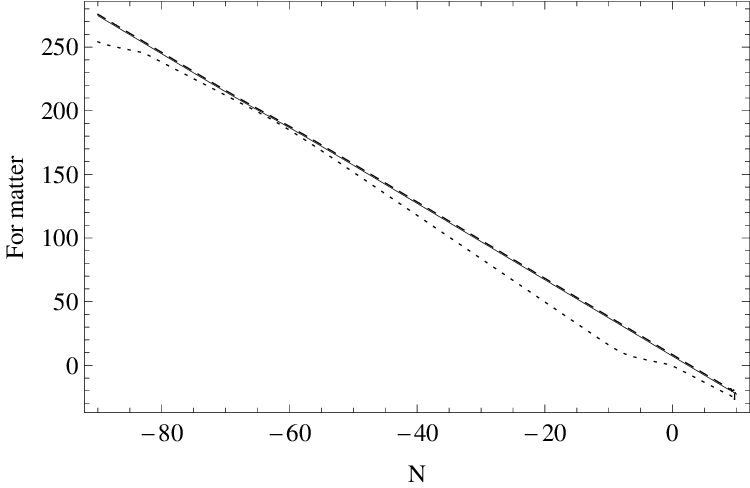}
\includegraphics[width=6cm]{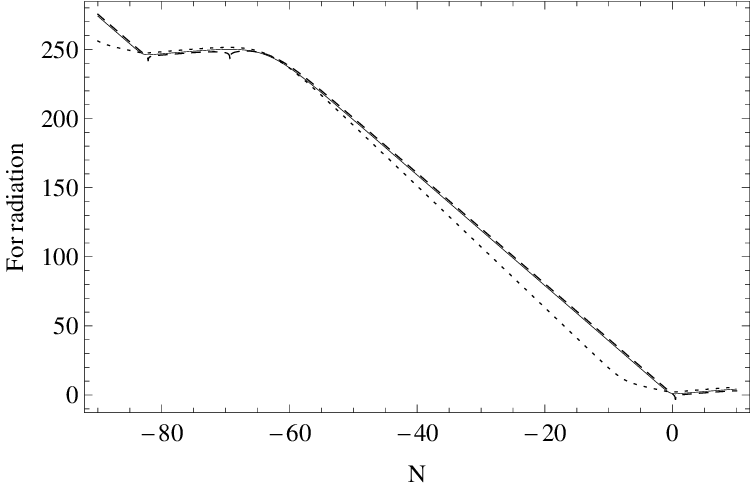}
\includegraphics[width=6cm]{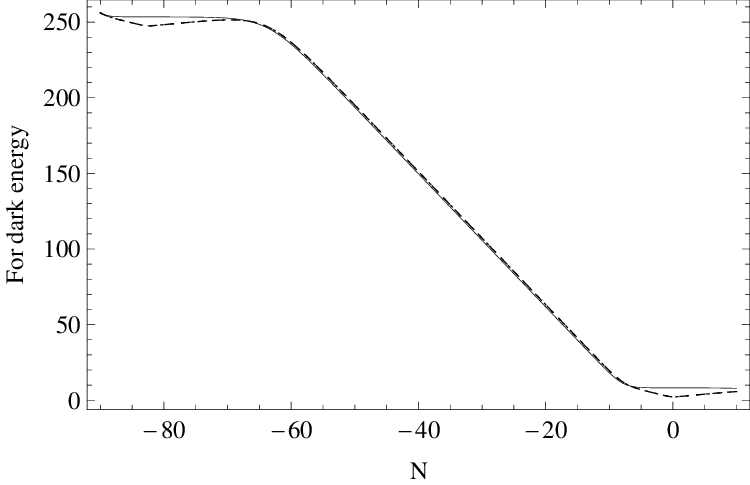}
\caption{\scriptsize{\label{fig6}The three above graphs show respectively a comparison between $\log \rho$ (thin), $\log \rho'$ (dashed) and the logarithm of the corresponding coupling function(s) divided by $H$ (dotted) for matter (first graph), radiation (second graph) and dark energy (third graph).}}
\end{figure}

\begin{figure}[h]
\centering
\includegraphics[width=6cm]{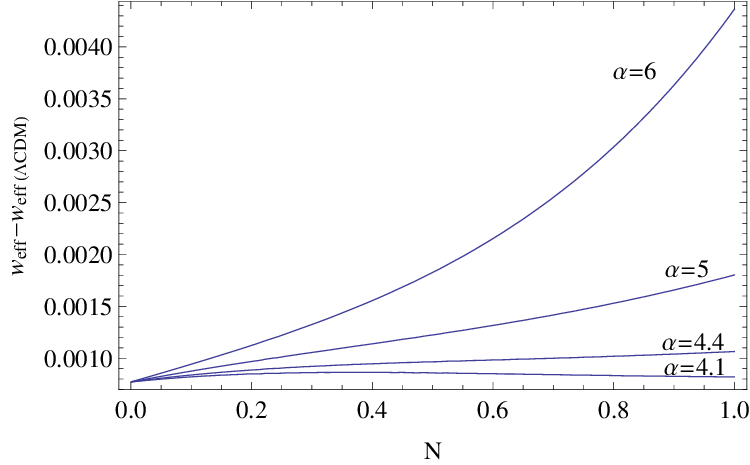}
\includegraphics[width=6cm]{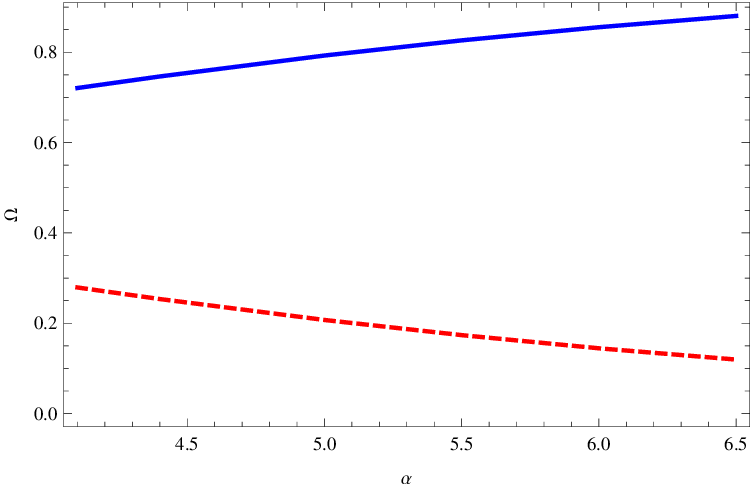}
\includegraphics[width=6cm]{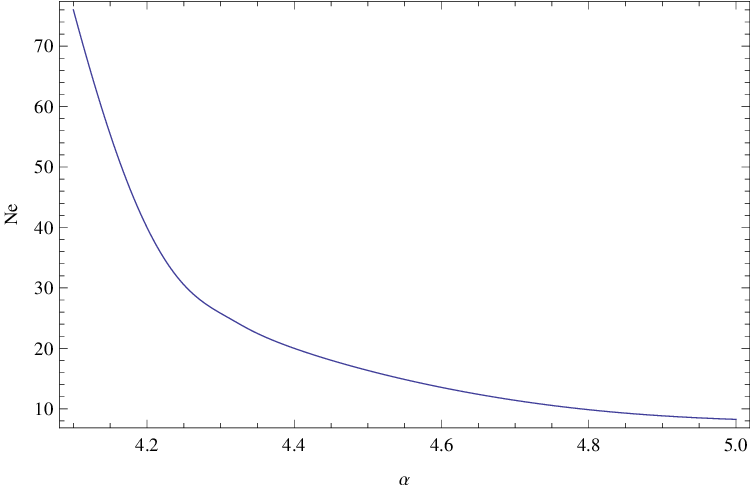}
\caption{\scriptsize{\label{fig8}First graph: evolution of $w_{eff}-w_{eff(\Lambda CDM)}$ with $\alpha$. Second graph: evolution of $\Omega_m$ (thin) and $\Omega_r$ (dashed) when $z=1080$ with $\alpha$. Third graph: evolution of $N_e$ with $\alpha$}}
\end{figure}

\begin{figure}[h]
\centering
\includegraphics[width=6cm]{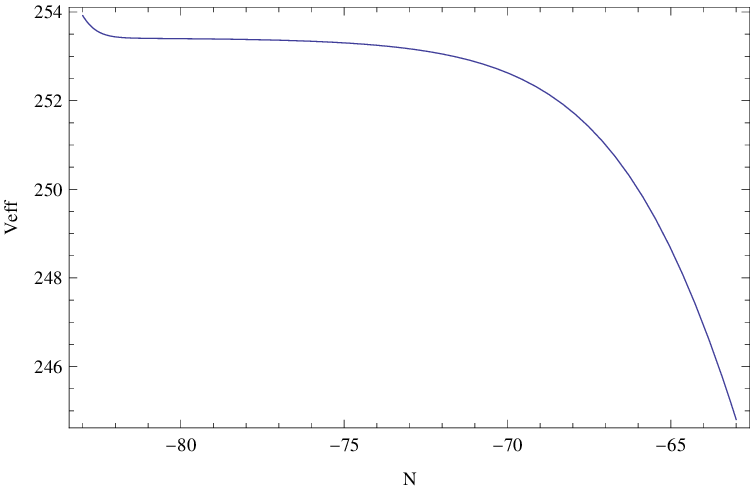}
\caption{\scriptsize{\label{fig9}Effective scalar field potential during the last inflation phase with $\alpha=4.4$, $H_0=70$, $\Omega_{m0}=0.27$ and $\Omega_{r0}=5.78\times 10^{-4}$.}}
\end{figure}

Now we look at one of the simplest forms of $q_m$ and $q_r$ respecting the constraints defined in the two previous subsections. We choose to consider $q_m=\Omega_m\Omega_r\Omega_d$ and $q_r=\alpha\Omega_r\Omega_d$. We then have $q_r/q_m=\alpha\Omega_m^{-1}$ that can be as large as necessary to satisfy the condition (\ref{ineq2}) of subsection \ref{s21} and allows Universe to approach a radiation dominated epoch after inflation. We also have $q_m$ and $q_r<<1$ along the trajectory followed by the standard $\Lambda CDM$ model in agreement with observations, as suggested in subsection \ref{s22}.\\\\
The equilibrium points of this cosmological model are
\begin{itemize}
\item $(y_1,y_2)=(0,1)$ with eigenvalues $(4-\alpha,1)$. It corresponds to the radiation dominated epoch. We assume it is a saddle and thus take $\alpha>4$ .
\item $(y_1,y_2)=(1,0)$ with eigenvalues $(-1,3)$. It corresponds to the matter dominated epoch and is a saddle.
\item $(y_1,y_2)=(0,0)$ with eigenvalues $(-3,-4+\alpha)$. It corresponds to the vacuum energy dominated epoch and is a saddle since we choose $\alpha>4$ .
\item $(y_1,y_2)=(\frac{(\alpha-4) \left(1-\alpha+\sqrt{-11+2 \alpha+\alpha^2}\right)}{2 (-3+\alpha)},\frac{1}{2} \left(1+\alpha-\sqrt{-11+2 \alpha+\alpha^2}\right)$. The related eigenvalues are
$$
\pm\sqrt{-\frac{(-4+\alpha) \left[\sqrt{-11+2 \alpha+\alpha^2} \left(-6+\alpha+\alpha^2\right)+\alpha \left(-11+2 \alpha+\alpha^2\right)\right]}{2(\alpha-3)}}
$$
When $\alpha>4$, these are purely imaginary numbers and this equilibrium point is thus a center(see \cite{Boy10} for instance).
\end{itemize}
There is no equilibrium sink point with $\alpha>4$. The trajectories of the phase space are homoclinic orbits (i.e. the orbits are closed curves) as shown on figure \ref{fig2}. To choose some values of $\alpha$, $H_0$, $\Omega_{m0}$ and $\Omega_{r0}$ in agreement with observations, we use the supernovae of the Union data set\cite{Ama10} and we require that at the CMB redshift $z=1080$, we recover the $\Lambda CDM$ values $\Omega_m (z=1080)\simeq 0.75$ and $\Omega_r (z=1080)\simeq 0.25$. We then calculate the best fit by minimising the following $\chi^2$
$$
\chi^2=\sum_{i=1}^{n}\frac{(m^{{obs}}_i-m_i^{{th}})^2}{\sigma^2_i}+\frac{(0.75-\Omega_m (z=1080))^2}{0.1^2}+\frac{(0.25-\Omega_r (z=1080))^2}{0.1^2}
$$
where $n$ is the number of data points, $m_i^{{\rm obs}}$ and $m_i^{{\rm th}}$ respectively the observed and theoretical magnitudes of supernovae, $\sigma_i$ the error bars and the two last terms, the priors at the CMB redshift. On figure \ref{fig7}, we plot the variation of $\chi^2$ with respect to $H_0$, $\Omega_{m0}$, $\Omega_{r0}$ and $\alpha$. Remark that this last parameter is not degenerated i.e., $\chi^2$ clearly varies with $\alpha$. The coupled model best fits the supernovae data as well as the $\Lambda CDM$ model when $\alpha=4.4$, $H_0=70$, $\Omega_{m0}=0.27$ and $\Omega_{r0}=5.78\times 10^{-4}$. Remark that this value of the radiation density parameter is larger than the value predicted by the $\Lambda CDM$ model. The corresponding trajectory in the phase space is shown as a thick trajectory on figure \ref{fig2} and is the one we consider in the rest of this section. \\\\
To study the above model, we define an "epoch" as the period of time between two moments when the $\Omega_i$ recover the same values. We also define without loss of generality, the beginning and thus the end of an epoch when $\Omega_r$ reaches its smallest value before increasing. In the $N$ time and for the above mentioned trajectory, an epoch then lasts $N_T=82.6$. During such an epoch, we have the following phases. At its beginning, expansion is accelerated, vacuum energy dominates and is cast into radiation whose parameter density $\Omega_r$ increases quickly. This can be considered as an inflation phase. When it ends, expansion decelerates and Universe goes to a radiation dominated epoch ($\Omega_r\simeq 1$). Then, the radiation starts to decrease and the expansion to be approximated by a $\Lambda CDM$ expansion (see below). Consequently, matter becomes the dominated species and then vacuum energy. A new accelerated expansion phase arises (comparable to the one in which we are today). When the radiation density parameter $\Omega_r$ starts to increase again, the epoch ends and a new one starts. The same scheme lasting $N_T$ repeats endlessly in the past and the future. This cosmological model thus unified "late time" expansion acceleration and inflation that are respectively the beginning (until $\Omega_r'>0$) and the end (from $\Omega_r'>0$) of a same accelerated expansion period when the thick trajectory crosses the gray area of figure \ref{fig2}.\\\\
Considering now the whole Universe evolution, the behaviours of density parameters and corresponding densities are plotted respectively on the first and second graph on figure \ref{fig3}. They show periodic behaviours in agreement with homoclinic orbits of figure \ref{fig2}. With the Hubble constant $H_0=70$, the Hubble function allows to calculate that Universe age is $13.8 Gy$ in agreement with observations. Then in the past, when $N\rightarrow -\infty$ and the proper time tends to a constant, the densities diverge and we have a singularity. In the future, the densities vanish and still oscillate.\\\\
One recovers these oscillations in the Hubble function plotted on the first graph on figure \ref{fig4}. In agreement with the decreasing densities, it tends to vanish in the future. This will not prevent alternated phases of decelerated and accelerated expansion since it depends on the sign of $\frac{\dot H}{H^2}+1$ and figure \ref{fig2} shows that the trajectory periodically crosses the set of phase space points where expansion is accelerated. The Hubble function being such as $d^2N/dt^2=H'H<0$, $N_T$ is an increasing period of time with respect to the proper time $t$. Hence, the epoch in which we are ($-81<N<1.6$) has a duration of $37.5Gy$ whereas the previous epoch ($-163.6<N<-81$) lasted around $10^{-43}y$.\\\\
The second graph on figure \ref{fig4} shows the behaviour of the coupling functions that are also oscillating. They diverge in the past and vanish in the future. Hence, although $q_m$ and $q_r$ tend to vanish periodically when $\Omega_d\simeq 0$ or $\Omega_r\simeq 0$, this is not necessarily the case of the coupling functions $Q_m$ and $Q_r$ because of the presence of the Hubble function in the denominator of $q_m$ and $q_r$. Consequently, $\rho_d$ can vary as shown on the second graph of figure \ref{fig3}, even when $q_m$ or $q_r$ vanish. Moreover, this does not prevent expansion at each epoch to be well approximated by a $\Lambda CDM$ expansion when the phase space trajectory is closed from $\Omega_d=0$ or $\Omega_r=0$ as explained below. Hence, as shown on figure \ref{fig5}, most of the evolution of the Hubble function during our epoch ($-81<N<1.6$) can be approximated by the standard $\Lambda CDM$ expansion ($H_0=70$, $\Omega_{m0}=0.27$ in agreement with supernovae data). In the same way, most of the evolution of the Hubble function during the previous epoch ($-163.6<N<-81$) can be approximated by a $\Lambda CDM$ expansion with $H_0\propto 10^{55}$ and $\Omega_{m0}\simeq 0$. When the transition to a new epoch arises, expansion stops being approximated by a $\Lambda CDM$ model until Universe becomes radiation dominated ($\Omega_r\simeq 1$) again.\\
\\
To understand how the Universe can be described by a $\Lambda CDM$ expansion despite the variation of the coupling functions, we have compared numerically the coupling functions with the densities and their derivatives in each energy conservation equations. This is plotted on figure \ref{fig6} for our epoch. Hence, concerning matter, $Q_m$ is always negligible in the energy conservation equation for $\rho_m$ that thus nearly behaves as $e^{-3N}$ like for the $\Lambda CDM$ model. Concerning the radiation, $Q_r$ is dominating from the transition between two epochs to the radiation dominated phase ($\Omega_r\simeq 1$) and negligible otherwise. We then have $\rho_r\propto e^{-4N}$ from the radiation dominated phase to the end of an epoch as for a $\Lambda CDM$ model. Concerning the dark energy, $Q_m+Q_r$ is dominating when $\rho_d$ is negligible and has then no consequence on Universe expansion. When dark energy starts to dominates with respect to the other species, we have first $\rho_d'=-(Q_m+Q_r)H^{-1}<<\rho_d$. The variation of $\rho_d$ with respect to its amplitude is thus very small and $\rho_d$ is nearly a constant (this is the phase in which we are today). Then, $Q_m+Q_r$ increases, $\rho_d$ decreases producing radiation and a new epoch begins. Hence, despite the variation of the $Q_i$, most of an epoch when $\Omega_r\simeq 0$ or $\Omega_d\simeq 0$ is approximated by a $\Lambda CDM$ expansion because then either the coupling functions are negligible in the energy conservation equations or they dominate the behaviour of a species when its density is negligible and is thus unable to move the expansion apart from a $\Lambda CDM$ behaviour.\\\\
We now examine the influence of the $\alpha$ parameter. Defining an effective density as $\rho_{deff}=\rho_m+\rho_r+\rho_d$, its effective equation of state is $w_{eff}=-2/3H'H^{-1}$. A similar reasoning can be made for the $\Lambda CDM$ model, thus defining $w_{eff(\Lambda CDM)}$. Then, we plot on the first graph of figure \ref{fig8} the difference $w_{eff}-w_{eff\Lambda CDM}$ for some values of $\alpha=(4.1,4.4,5,6)$ and for the recent epoch $0<N<1$. We remark that $w_{eff}-w_{eff\Lambda CDM}$ is smaller than  $10^{-2}$. Such a difference is not observable today with available data. However, the figure \ref{fig7} clearly shows that $\chi^2$ varies slowly but in a detectable way with $\alpha$. This indicates that the variation of $\chi^2$ with $\alpha$ should be due to the priors. Thus the variation of $\alpha$ should be responsible for a variation of $\Omega_m$ and $\Omega_r$ at the CMB redshift $z=1080$. This is confirmed by the second graph on figure \ref{fig8}. The larger $\alpha$, the larger the matter and the smaller the radiation densities parameters in $z=1080$.\\
We also determined the influence of $\alpha$ on the number of e-fold $N_e$ during an inflation phase. When $\alpha=4.4$, we have $N_e\simeq 20$ as shown on the first graph (gray areas) of figure \ref{fig4}. This seems very few with respect to what is necessary to solve the flatness and horizon problems, i.e $N_e\simeq 60$. However, there is also an infinite number of such inflation phases with the same $N_e$ in the past (that take place during the infinite number of previous epochs) and this should be enough to solve the flatness and horizon problems. Nevertheless, we can study how $N_e$ varies with $\alpha$ and show that a small variation of $\alpha$ with respect to its best fitting value $4.4$ is enough to get $N_e\simeq 60$ already during the last inflation phase. This is illustrated on the third graph on figure \ref{fig8}. The smaller $\alpha$, the larger $N_e$. In particular, we have repeated inflation phases with $N_e\simeq 60$ when $\alpha\simeq 4.14$. This last value is still in very good agreement with supernovae data but would imply that $\Omega_m=0.72$ and $\Omega_r=0.28$ at the CMB redshift. These values are slightly different from the ones predicted by the $\Lambda CDM$ model. So, either we take $\alpha=4.4$, flatness and horizon problems are solved by several inflation phases and we recover the same values as the $\Lambda CDM$ model for $\Omega_m$ and $\Omega_r$ at the CMB redshift. Or we take $\alpha=4.14$, flatness and horizon problems are already solved by the last inflation phase but $\Omega_m$ and $\Omega_r$ at the CMB redshift have values slightly different from the ones predicted by the $\Lambda CDM$ model.
\\\\
Finally, let us write some few words about structure formation. A complete study is out of the scope of this paper so we just present some simple calculations. First, our model should be able to generate the large scale structures in the Universe as, for instance, scalar field inflation\cite{Lid99}. To check that, we consider the last inflation phase and we calculate the effective scalar field potential $V_{eff}$ that our coupled model mimics. Still defining the density of an effective fluid as the sum of all the densities, we rewrite equations (\ref{H2}-\ref{rhodd}) as 
$$
H^2=\frac{k}{3}\rho_{deff}
$$
with $\rho_{deff}=\rho_m+\rho_r+\rho_d$. Summing the energy conservation equation, we also get
$$
\dot \rho_{deff}+3H(1+w_{eff})\rho_{deff}=0
$$
where this time we write $w_{eff}$ as
$$
1+w_{eff}=\frac{3\rho_m+4\rho_r}{3\rho_{deff}}
$$
During the inflation phase $\rho_{deff}\simeq \rho_d$ and accelerates Universe expansion. To find the effective scalar field potential equivalent to $\rho_{deff}$, we then define the scalar field derivative and the effective potential as
$$
\dot\phi^2=\rho_{deff}+p_{deff}
$$
with $p_{deff}=w_{eff}\rho_{deff}$ and
$$
V_{eff}=(\rho_{deff}-p_{deff})/2
$$
All the quantities with the "$eff$" index can thus be written as some functions of $\rho_m$, $\rho_r$ and $\rho_d$ that we can obtain numerically. We still consider the best fitting case of our coupled model with $\alpha=4.4$ and then get the figure \ref{fig9} showing the effective scalar field potential during the last inflation phase. It is slowly varying during most part of the inflation period that ends with the potential decays. Calculating the slow roll parameters $(dV_{eff}/d\phi/V_{eff})^2$ and $d^2V_{eff}/d\phi^2/V_{eff}$, we find that they are well below unity when the potential is flat. All this is in agreement with the fact that as long as $\rho_d>>(\rho_m,\rho_r)$ during inflation, $w_{eff}\simeq -1$. The coupled model is thus equivalent during an inflation phase to the presence of a scalar field with a flat potential. It should thus be able to generate the large scale structures as the scalar field inflation.\\
Another question is then to know if the large scale structures generated by the coupled model can be in agreement with observations. To answer this question, we calculate the matter density contrast $\delta_m$ for the coupled model and compare it with the one of the $\Lambda CDM$ model when the redshift $z$ is in the range $0<z<2$. Here, we define an effective dark energy as $\rho_{deff}=\rho_d+\rho_r$. This allows to rewrite equations (\ref{H2}-\ref{rhodd}) as 
$$
H^2=\frac{k}{3}(\rho_m+\rho_{deff})
$$
$$
\dot \rho_m+3H\rho_m=Q_m
$$
$$
\dot\rho_{deff}+3(1+w_{eff})H\rho_{deff}=-Q_m
$$
with $1+w_{eff}=\frac{4}{3}\rho_r/\rho_{deff}\not = 0$. Under this form, only the radiation terms have been absorbed in the effective dark energy terms. $w_{eff}$ now describes an effective dark energy different from vacuum that is coupled to dark matter. This allows to calculate $\delta_m$ that writes in this case\cite{Ame04,Bor08,Dev15}
\begin{equation}\label{pert}
\ddot\delta_m+\dot\delta_m(2 H+\frac{Q_m}{\rho_m})+\delta_m(-\frac{3}{2}\rho_m+2\frac{ H Q_m}{\rho_m}-\frac{Q_m\dot{\rho_m}}{\rho_m^2}+\frac{\dot{Q_m}}{\rho_m})=0
\end{equation}
Comparing $\delta_m$ for the coupled model to the one, $\delta_{m(\Lambda CDM)}$, of the $\Lambda CDM$ model when $\Omega_{m0}=0.27$ and $H_0=70$, we find that $\delta_m-\delta_{m(\Lambda CDM)}<10^{-3}$ whatever $\alpha$ that we tested in the interval $4.1<\alpha<10$. This means that the predictions about structures formation for the coupled model and the $\Lambda CDM$ model both fit observational data as well and are indistinguishable with present day observations whose error bars are of the order of $10^{-1}$\cite{Tad14,Ame14}.
\\\\
A concluding remark is that the epoch in which we live is the first one that lasts long enough to allow life emergence. While it could appear as a coincidence, it is interesting to note that it is in agreement with recent results from \cite{BehPee15} estimating that Universe will form over $10$ times more planets than currently exist.
\section{Conclusion} \label{s4}
In this paper, we first look for some necessary conditions on the coupling functions such as Universe exits from inflation to go to a radiation dominated epoch in $\Omega_{m(dom)}<<1$ and $\Omega_{r(dom)}\simeq 1$ and then that its expansion could behave as the one of a $\Lambda CDM$ model in agreement with our present day Universe. We have shown that the first requirement about inflation to radiation transition implies the necessary condition $q_r/q_m>(\Omega_{r(dom)}-3/4)\Omega_{m(dom)}^{-1}$. The second requirement about the $\Lambda CDM$ model implies the necessary condition that after Universe reaches the point $(\Omega_{m(dom)},\Omega_{r(dom)})$, $q_m$ and $q_r<<1$ near the phase space lines $\Omega_d=0$ and $\Omega_r=0$. Then one of the simplest forms of coupling functions $Q_m$ and $Q_r$ respecting these two requirements is defined by $q_m=\Omega_m\Omega_r\Omega_d$ and $q_r=\alpha\Omega_r\Omega_d$.\\
When $\alpha>4$, this model describes a Universe starting with a singularity and evolving with endless alternated phases of accelerated and decelerated expansion. It thus unifies inflation and late time acceleration in a same physical phenomenon, repeating periodically. Such an oscillating behaviour has been considered in \cite{Dod00} to help to solve the coincidence problem with an oscillating (uncoupled) scalar field potential. It has also been studied in dark fluid with archimedean-type force where dark matter interact with dark energy\cite{Bal11}, in quintom cosmology\cite{Min14} or in inhomogeneous plane symmetric space-time\cite{She14}. We have shown that the values $\alpha=4.4$, $\Omega_{m0}=0.27$, $\Omega_{r0}=5.78\times 10^{-4}$ and $H_0=70$ are in agreement with supernovae data and the parameter densities values of the $\Lambda CDM$ model at the CMB redshift $z=1080$. The number of e-fold during the last inflation phase is then around $20$. This is small but there was an infinite number of inflation phases with the same number of e-fold before it. Moreover, this number can be increased to $60$ by considering slightly smaller values of $\alpha$ if necessary. It then changes a little bit the values of $\Omega_m$ and $\Omega_r$ at the CMB redshift with respect to what is predicted by the $\Lambda CDM$ model and is still in agreement with the supernovae data. We have given some arguments seeming to show that formation structures should take place as with a scalar field inflation and that the matter density contrast predicted by the coupled model cannot be distinguished from the one of the $\Lambda CDM$ model. The Universe age is similar to this predicted by the $\Lambda CDM$ model. Moreover, for most of the time of an epoch when $\Omega_r\simeq 0$ or $\Omega_d\simeq 0$, expansion can be approximated by a $\Lambda CDM$ expansion, the vacuum energy varying strongly when it is negligible before being nearly a constant when it becomes the dominating species. We would be now ($N=0$) entering in a phase approaching a De Sitter behaviour (i.e. $H\simeq const.$) and that would last around $23Gy$. After that time, the vacuum would decay again into radiation and Universe would experiment a new inflation phase with smaller and smaller energy densities.
\bibliographystyle{unsrt}

\begin{thebibliography}{10}
\bibitem{Car01}
S. M. Carroll,
\newblock Living Rev. Relativity, 4, (2001).

\bibitem{Lim13}
J. A. S. Lima et al.,
\newblock MNRAS, Vol.437 3331-3342 (2013).

\bibitem{Per99}
S. Perlmutter et al., 
\newblock Astrophys. J., 517, 565-586 (1999).

\bibitem{Rie98}
A. G. Riess et al, 
\newblock  Astron. J., 116, 1009-1038 (1998).

\bibitem{Gar93}
J. Garcia-Bellido, 
\newblock Int.J.Mod.Phys.D2:85-95 (1993).

\bibitem{Wet95}
C. Wetterich, 
\newblock Astron.Astrophys.301:321-328 (1995).

\bibitem{Ame00}
L. Amendola., 
\newblock MNRAS, 312:521 (2000).

\bibitem{Oli05}
G. Olivares et al, 
\newblock Phys.Rev. D71, 063523 (2005).

\bibitem{Yan14}
W. Yang \& L. Xu, 
\newblock Phys. Rev. D 89, 083517 (2014).

\bibitem{Cos14}
A. A. Costa et al, 
\newblock Phys. Rev. D 89, 103531 (2014).

\bibitem{Gas87}
M. Gasperini, 
\newblock Phys.Lett. B194, 347 (1987).

\bibitem{Pav88}
D. Pavon, 
\newblock Phys. Lett. B215,1, (1988).

\bibitem{Lim96}
J. A. S. Lima,
\newblock Phys.Rev. D54 2571-2577 (1996).

\bibitem{Bar06}
G. Barenboim \& J. D. Lykken,
\newblock JHEP 0607:016 (2006).

\bibitem{Hor07}
R. Horvat \& D. Pavon, 
\newblock Phys. Lett. B653:373-377 (2007).

\bibitem{Suy08}
T. Suyama \& M. Yamaguchi,
\newblock Phys.Rev.D77:023505 (2008).

\bibitem{Her10}
R. Herrera,
\newblock Phys.Rev.D81:123511 (2010).

\bibitem{Gut81}
A. Guth,
\newblock Phys. Rev. D23, 2, 347 (1981). 

\bibitem{Bra01}
R. H. Brandenberger,
\newblock proceedings BROWN-HET-1256 (2001). 

\bibitem{Lin04}
A. Linde,
\newblock Phys.Scripta T117:40-48 (2005).

\bibitem{Bic14}
The BICEP2 Collaboration, Phys. Rev. Lett. 112, 241101 (2014).

\bibitem{Ada14}
R. Adam et al.,
\newblock A$\&$A, 586, A133 (2016).

\bibitem{Per13}
E. L. D. Perico et al,
\newblock Phys. Rev. D 88, 063531 (2013).

\bibitem{Lim15}
J. A. S. Lima et al.,
\newblock Gen.Rel.Grav. 47, 40 (2015).

\bibitem{Bea01}
R. Bean \& al, 
\newblock Phys. Rev. D64:103508 (2001).

\bibitem{Fre87}
K. Freese \& al,
\newblock Nucl. Phys. B, Volume 287, p. 797-814 (1987).

\bibitem{Cam15}
S. d. Campo et al,
\newblock Phys. Rev. D91, 123539 (2015).

\bibitem{Mur16}
R. Murgia \& al,
\newblock JCAP 04,014 (2016).

\bibitem{Boe08}
C. G. Boehmer \& al,
\newblock Phys.Rev.D78:023505 (2008).

\bibitem{Oli06}
G. Olivares\& al,
\newblock Phys.Rev.D74:043521 (2006).

\bibitem{Boh10}
C. G. Boehmer et al., 
\newblock Phys.Rev.D81:083003 (2010).

\bibitem{Ama10}
Amanullah et al.,
\newblock {\em Ap.J.}, 716:712-738, 2010.

\bibitem{Boy10}
J. R. Brannan \& W. E. Boyce,
Differential Equations: An Introduction to Modern Methods and Applications
\newblock {\em Wiley}, 2nd editio,, chapter 3.4, p177, (2010).

\bibitem{Lid99}
A. R. Liddle,
\newblock astro-ph/9901124.
\newblock {\em proceedings of ICTP summer school in high-energy physics}, 15, (1999).

\bibitem{Ame04}
L. Amendola., 
\newblock Phys.Rev.D69:103524, (2004).

\bibitem{Bor08}
H. A. Borges et al, 
\newblock Phys.Rev.D77:043513, (2008).

\bibitem{Dev15}
N. C. Devi et al, 
\newblock MNRAS, 448, 37-41, (2015).

\bibitem{Ame14}
L. Amendola et al, 
\newblock Phys. Rev. D89, 063538, (2014).

\bibitem{Tad14}
L. Taddei et L. Amendola, 
\newblock arXiv:1408.3520,(2014).

\bibitem{BehPee15}
P. Behroozi \& M. Peeples 
\newblock {\em MNRAS}, 454, 2, 2015.

\bibitem{Dod00}
S. Dodelson \& al,
\newblock Phys.Rev.Lett.85:5276-5279 (2000).

\bibitem{Bal11}
A. B. Balakin \& V. V. Bochkarev,
\newblock Phys.Rev.D83:024036 (2011).

\bibitem{Min14}
S. Ming \& Z. Liang,
\newblock Chinese Physics Letters, 31, 1, (2014).

\bibitem{She14}
M. She \& L. P. Jiang,
\newblock Astrophysics and Space Science, 354, 2, (2014).

\end{thebibliography}

\end{document}